\documentclass[12pt]{iopart}

\usepackage{iopams}  
\usepackage{float}
\usepackage{graphicx}
\usepackage{tabularx}
\usepackage{multirow}

\begin{document}

\title[Iterative symphonic tunneling for satisfiability problems]{Iterative quantum optimization of spin glass problems with rapidly oscillating transverse fields}

\author{Brandon Barton$^{1,2,3\dagger}$, Jacob Sagal$^{1\dagger}$, Sean Feeney$^{1}$, George Grattan$^{1,5}$, Pratik Patnaik$^{4}$, Vadim Oganesyan$^{3,6,7}$, Lincoln D Carr$^{1,2,4}$ and Eliot Kapit$^{1,4}$}

\address{$^1$ Quantum Engineering Program, Colorado School of Mines, 1523 Illinois St, Golden, CO 80401, United States of America}
\address{$^2$ Department of Applied Mathematics and Statistics, Colorado School of Mines, 1500 Illinois St, Golden, CO 80401, United States of America}
\address{$^3$ Center for Computational Quantum Physics, Flatiron Institute, 162 5th Avenue, New York, NY 10010, USA}
\address{$^4$ Department of Physics, Colorado School of Mines, 1523 Illinois St, Golden, CO 80401, United States of America}
\address{$^5$ Department of Computer Science, Colorado School of Mines, 1500 Illinois St, Golden CO 80401, United States of America}
\address{$^6$ Department of Physics and Astronomy, College of Staten Island, City University of New York, Staten Island, NY 10314, United States of America}
\address{$^7$ Physics program and Initiative for the Theoretical Sciences, The Graduate Center, City University of New York, New York, NY 10016, United States of America}

\ead{ekapit@mines.edu}
\vspace{10pt}

\begin{indented}
\item[]July 2024
\end{indented}

\begin{abstract}
    In this work, we introduce a new iterative quantum algorithm, called Iterative Symphonic Tunneling for Satisfiability problems (IST-SAT), which solves quantum spin glass optimization problems using high-frequency oscillating transverse fields. IST-SAT operates as a sequence of iterations, in which bitstrings returned from one iteration are used to set spin-dependent phases in oscillating transverse fields in the next iteration. Over several iterations, the novel mechanism of the algorithm steers the system toward the problem ground state. We benchmark IST-SAT on sets of hard MAX-3-XORSAT problem instances with exact state vector simulation, and report polynomial speedups over trotterized adiabatic quantum computation (TAQC) and the best known semi-greedy classical algorithm. When IST-SAT is seeded with a sufficiently good initial approximation, the algorithm converges to exact solution(s) in a polynomial number of iterations. Our numerical results identify a critial Hamming radius(CHR), or quality of initial approximation, where the time-to-solution crosses from exponential to polynomial scaling in problem size. By combining IST-SAT with future classical or quantum approximation algorithms, larger gains may be achieved. The mechanism we present in this work thus presents a new path toward achieving quantum advantage in optimization.
\end{abstract}

\vspace{2pc}

\section{Introduction}
Binary constraint satisfaction problems represent a promising opportunity to achieve practical quantum advantage in real-world problems found in optimization, artificial intelligence, and cryptography. In the worst case, and often, the typical case, these problems are exponentially hard to solve or even approximate for classical machines. Therefore, an efficient solution mechanism would have broad impacts and applicability.

So far however, demonstrating consistent quantum advantage with heuristic algorithms has remained elusive. A wide array of heuristic quantum algorithms such as analog quantum annealing \cite{finnila1994quantum,kadowakinishimori1998,das2008colloquium,johnson2011quantum,boixo2014evidence,hauke2019perspectives,king2023quantum}, adiabatic quantum computing (AQC) \cite{farhigoldstone2000,RevModPhys.90.015002}, quantum approximation optimization algorithms (QAOA) \cite{farhi2014quantum}, and more exotic variations such as adaptive derivative assembled problem tailored (ADAPT)-QAOA \cite{zhu2022adaptive}, recursive QAOA \cite{bravyi2022hybrid}, QAOA supplemented with amplitude amplification \cite{doi:10.1126/sciadv.adm6761}, and energy matching/population transfer algorithms \cite{baldwin2018quantum,PhysRevX.10.011017,kechedzhi2018efficient,smelyanskiy2019intermittency}, have been proposed for spin glass optimization. Their empirical performance, however, has been decidedly mixed, and the frequently observed quadratic speedups from schedule optimization (a scheme dating back to the adiabatic formulation of Grover's search \cite{rolandcerf2002}) are fragile and likely not feasible at large $N$ \cite{kapit2021noise}. Furthermore, if the problem Hamiltonian is stoquastic \cite{bravyidivincenzo2006}, quantum Monte Carlo \cite{isakov2016understanding,andriyash2017can,jiang2017scaling,jiang2017path,king2019scaling} can simulate small systems sizes in the quadratic (incoherent) scaling limit. 

Motivated by the recent search for applications of quantum algorithms to machine learning, variational quantum algorithms have also also been applied to spin glass optimization problems \cite{cerezo2021variational}. Yet, large gains from these methods may be limited by the cost of computing gradients and may often get stuck in local minima and encounter barren plateaus, or loss of variance in the optimization landscape \cite{ragone2023unified,Wang_2021}. Thus, the full capabilities of quantum computing in this space, and potential asymptotic limits for very general algorithms in this class, deserve further exploration. Given that beyond-quadratic speedups are critical to achieve useful quantum advantage \cite{babbush2021focus}, new mechanisms that expand the quantum optimization toolbox for reaching exact solutions \cite{arora2009computational}, or even approximate solutions may have significant impacts.

In this work, we propose a non-classical steering mechanism that guides quantum optimization algorithms towards the true ground state in spin glass problems. We demonstrate this mechanism by introducing a new heuristic quantum algorithm which we call \emph{Iterative Symphonic Tunneling for Satisfiability problems} (IST-SAT). The IST-SAT algorithm modifies quasi-continuous time Trotterized AQC (TAQC), with total evolution time increasing linearly with 
$N$, by adding a monochromatic fast oscillating field along $Y$ to all spins. The addition of the oscillating field is inspired by a recently discovered mechanism termed \emph{symphonic tunneling} which has exponentially accelerated macroscopic quantum tunneling rates between ferromagnetic ground states  \cite{mossi2023embedding, grattan2024exponential}. IST-SAT builds on this work by expanding the use of high frequency oscillating transverse fields to quantum optimization. We consider disordered spin glass instances drawn from the MAX-3-XORSAT problem class \cite{arora2009computational}, where the problem Hamiltonian $H_P$ is diagonal in $Z$ and the DC transverse field is along $X$. In contrast to accelerating collective tunneling in the transverse-field Ising model (TFIM) where the classical ground states are known a priori and there is no frustration \cite{grattan2024exponential}, applying a uniform oscillating field with identical frequencies and phases to all sites produces no meaningful benefits for finding low energy states of MAX-3-XORSAT problems.
 
 To formulate the IST-SAT algorithm, we propose an iterative strategy, where bitstrings returned from TAQC 
 are used to set spin-dependent phases for the oscillating drive terms in subsequent iterations of the algorithm. 
 We define the list of
 phases $P = \left[ \varphi_1, \varphi_2, \dots , \varphi_N \right]$, with $\varphi_j 
\in\{0,\pi\}$ assigned to each spin.
We test the performance of IST-SAT on the planted partial solution problem (PPSP) class defined in \cite{kapit2024approximability}, and with extensive exact state-vector simulations. Our results show that when \emph{the phase pattern} more closely matches the pattern of bits in the planted classical ground state $G$, the probabilities of reaching $G$ and states nearby to the global optima increase monotonically. Our simulations identify a \emph{critical Hamming radius} (CHR) $r_c$ where the time to find $G$ in iterative optimization scales polynomially, starting from any bitstring with Hamming distance $D_\mathrm{H} \leq r_c N$ from $G$. IST-SAT achieves a polynomial speedup over both trotterized AQC (TAQC) and the best-known quasi-greedy classical algorithm for this problem. We also demonstrate that using TAQC as a seed algorithm for classical methods yields meaningful polynomial speedups over both algorithms on their own. Taken together, these results suggest new and promising routes to gradient-free quantum optimization, through a new non-classical steering mechanism to find ground-states with quantum algorithms.

\section{Problem and algorithm definitions}\label{probdef}

In contrast to many problems where asymptotic exponential scaling is not reached until prohibitively large system sizes for classical computers and NISQ era devices \cite{mohseni2022ising}, 3-XORSAT problems have exponential scaling that is typically observed at small $N$. We test IST-SAT on the MAX-3-XORSAT \cite{arora2009computational} problem, defined on a random 3-uniform hypergraph consisting of $N_C$ three-body constraint terms. The problem Hamiltonian is given by
\begin{eqnarray}
    H_P = -\sum_{ijk}^{N_C} V_{ijk} Z_i Z_j Z_k, \; \; V_{ijk} = \pm 1,
\end{eqnarray}
where for a given bitstring a constraint is satisfied if $\langle V_{ijk} Z_i Z_j Z_k\rangle = +1$, and unsatisfied otherwise. Since the problem is linear, Gaussian elimination can be used to check if the problem is satisfiable in polynomial time. However, when not all the constraints can be satisfied, finding the lowest energy state(s) is known to be NP-hard \cite{haastad2001some}.

This work utilizes a family of instances called planted partial solution problems (PPSPs), used in recent quantum approximation algorithms for MAX-3-XORSAT \cite{kapit2024approximability}. To construct a PPSP we first pick a random hypergraph with $N$ variables that participate in $N_C$ constraints with $N_C \gg N$, a fixed small fraction $\epsilon$ (we use $\epsilon=0.1$ in this work), and a random bitstring $G$ to be the planted ground state. We randomly select $\of{1-\epsilon} N_C$ constraints to be satisfied in $G$ (by choosing the signs of the $V_{ijk}$) with the remaining left unsatisfied. For small $\epsilon$ and $N_C \gg N$, this construction makes $G$ a unique ground state with high probability (the SAT/UNSAT transition here is at $N_C/N \sim 0.92$ \cite{dubois20023}). We note that the PPSP instances in this work can be more difficult than 3-regular hypergraphs studied in previous work \cite{bellitti2021entropic,kowalsky20213}, which are easy to approximate and can be solved efficiently if satisfiable. Throughout this work, we used the set of constraint densities $N_C/N \in \cuof{1.5,2,4}$ to test the performance of IST-SAT over several densities above the SAT/UNSAT transition.

\begin{figure}
    \centering
    \includegraphics[width=0.99\textwidth]{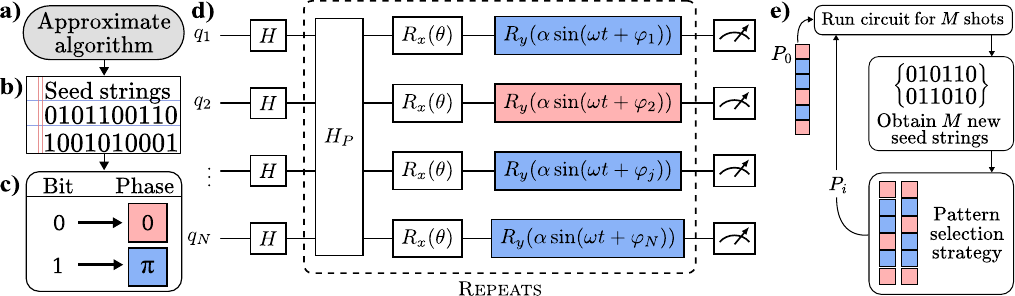}
    \caption{Schematic of the IST-SAT algorithm. An approximate algorithm--which can include random guessing and high-depth TAQC--in (a) is used to find ``seed strings" (b) for the phase pattern in the $H_\mathrm{ST}$ drive Hamiltonian. The local phases $\{ \varphi_1, \varphi_2, \dots, \varphi_N \}$ in $H_\mathrm{ST}$ are set from bits of the seed states using the key (c), which we denote with colored $R_{y}$ rotations in (d). After providing the algorithm an initial phase pattern $P_0$, the iterative process described in (e) continues, where the circuit is run for $M$ shots, generating new seed strings which are converted to phase patterns. A new phase pattern $P_i$ is then selected to be used in the next iteration. The total number of "REPEATS" is set by the ratio of total runtime and Trotter step, $t/dt$ (see text).}\label{ISTfig}
\end{figure}

IST-SAT starts from the ``standard" quasi-continuous time AQC method of interpolating between a transverse field ``driver" Hamiltonian and the problem Hamiltonian with Trotterized evolution from $t=0$ to $t=t_f$:
\begin{eqnarray}\label{Hbase}
    H(t) &=& f(t) H_{D} + g(t) H_{P}, \; \; H_{D} = - \sum_j X_j, \\
    f(0) &=& g(t_f) = 1, \; \;  f(t_f) = g(0) = 0. \nonumber
\end{eqnarray}
We interpolate between the problem and driver Hamiltonians using the following functions
\begin{eqnarray}\label{profiles}
f(t) = \sqrt{1- t/t_f}, \; \; g(t) = \sqrt{t/t_f},
\end{eqnarray}
The choices of $f(t)$ and $g(t)$ empirically outperform linear interpolation, producing a better prefactor and modestly better scaling with $N$, though we expect the asymptotic scaling of the two schedules may converge for this problem. We let the total evolution time $t_f$ grow linearly with $N$. 

IST-SAT modifies the base Hamiltonian in equation \ref{Hbase} by adding a high-frequency monochromatic oscillating field 
\begin{equation}
    H_{\mathrm{ST}}(t) = \alpha \sum_j Y_j \sin(\omega t + \varphi_j).
\end{equation}
Within $H_{\mathrm{ST}}(t)$, the parameters $\varphi_j = \cuof{0,\pi}$ are single-qubit phases, and $\alpha = \alpha_s \ln N$ is the drive strength. The total Hamiltonian is then given by 
\begin{eqnarray}\label{HIST}
    H(t) = f(t) H_{D} + g(t) H_{P} + h(t) H_{ST}(t),
\end{eqnarray}
where $h(t)$ is a smooth function with initial and final conditions $h(t=0),h(t_f) = 0$.  In this work, we used $h(t) = 4 \sqrt{(1-t/t_f)(t/t_f)}$, $\alpha_s = 0.6$ and $\omega = 2\pi \times 6 \ln N$. The choice of $\alpha$ and $\omega$ both increasing logarithmically with $N$ minimizes heating \cite{abanin2015exponentially,kuwahara2016floquet} while ensuring that the novel low-frequency terms generated by the high frequency drive have constant magnitude. Time evolution is implemented via Trotterization and gate-based exact state-vector simulation using the qulacs python package \cite{suzuki2021qulacs}.

The schematic work-flow of the IST-SAT algorithm is shown in Fig. \ref{ISTfig}, which begins with an approximate algorithm. The initial approximation algorithm can be a greedy classical approach, simulated annealing, or a quantum algorithm such as AQC/TAQC or QAOA. The initial approximate algorithm is used to obtain \emph{seed strings} which are used to set phases in the high frequency AC drive $H_{\mathrm{ST}}(t)$. We note the method of phase selection in $H_{\mathrm{ST}}$ is distinct from \emph{warm start} methods \cite{cain2023qaoa}, where preparing a good initial state with high overlap to the ground state is costly, and still expected to be hard. Instead, IST-SAT starts from the initial superposition state $\vert + \rangle^{\otimes N}$, and sets parameters in the time-dependent Hamiltonian $H_{\mathrm{ST}}(t)$. In one iteration of IST-SAT, the binary configuration of a selected seed string is used to form the phase pattern $P$, which sets parametrizes the quantum circuit in figure \ref{ISTfig}. After evolving the circuit, new measurements in the $z$-basis generate $M$ new phase patterns that may be used in further iterations.

For the time evolution of equation \ref{HIST}, we used a mean evolution time of $t_f = N/32$, with $dt = 0.4/\omega$ to ensure high frequencies are appropriately sampled. The total circuit depth thus scales as $N \ln N$ problem Hamiltonian applications. All data is averaged over 1000 random problem instances for each $N$. The total run-time is total averaged between $T =2 t_f/3$ and $T =4 t_f/3$, for all simulations of IST-SAT and TAQC. This averaging procedure has shown to effectively smooth out unpredictable diabatic effects that can make reliably estimating scaling difficult \cite{mossi2023embedding}, particularly when success probabilities decay exponentially with $N$. The parameters used in this work are the result of trial and error on small system sizes. Therefore further parameter optimization techniques are expected to produce further benefits.

\section{Performance results of IST-SAT}\label{sec:results}

\begin{figure}
    \centering
    \includegraphics[width=\textwidth]{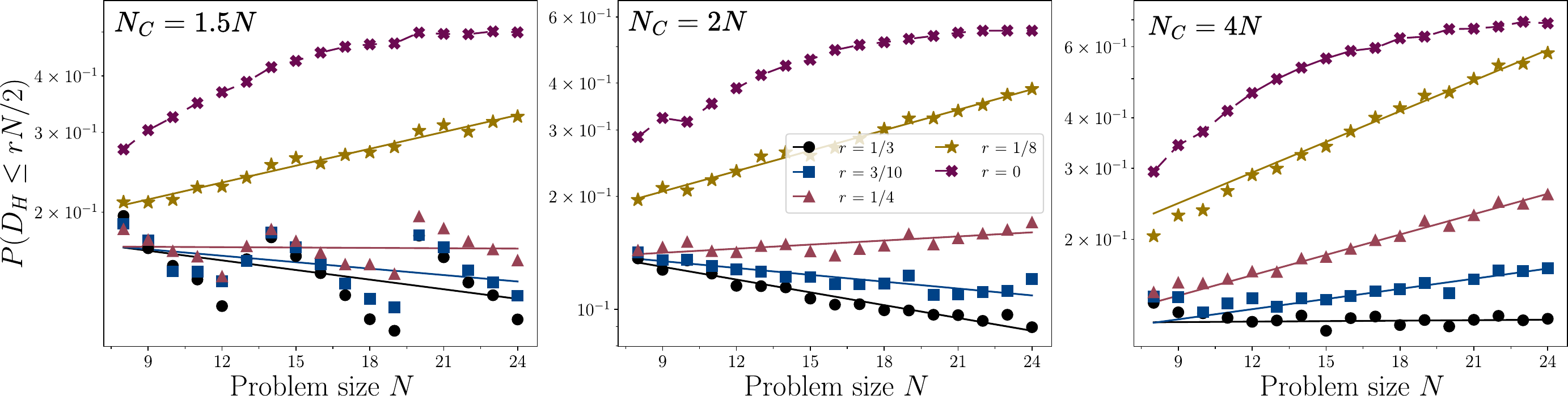}
    \caption{Exponential to polynomial scaling transition for IST-SAT seeded with random initial phase patterns $P_0$, with fraction $(1-r)$ of the bits matching the planted solution. For constraint densities $N_C/N = {1.5, 2, 4}$ (left, middle, right), the plot shows the probability of returning a solution within $D_{\mathrm{H}} \leq r N/2$ flips of the nearest solution. For probability calculations with $D_{\mathrm{H}} \leq r N$, refer to \ref{app:further_istsat_details}.}
    \label{fig:ISTSAT_PrNby2_main_text}
\end{figure}

The results from state-vector simulation of IST-SAT shown in figure \ref{fig:ISTSAT_PrNby2_main_text}, demonstrate that an appropriately chosen phase pattern dramatically accelerates a collective rearrangement of spins at the transition from the initial paramagnetic state to ground states of the problem Hamiltonian. Of course, this choice of phase pattern is not obvious since solutions to the problem are generally not known, and the purpose of the algorithm is to find it. If the phases are chosen randomly, the protocol in equation \ref{HIST} shows no scaling advantage over the uniform field protocol in equation \ref{Hbase} with a modest prefactor disadvantage. Conversely, if we let $\varphi_j = \pi b_j$, where $b_j \in\cuof{0,1}$ is the value of bit $j$ in the planted solution $G$, $P_{\mathrm{GS}} \of{t_f}$ is empirically \emph{constant} with linear runtime $t_f \propto N$ (see the $r=0$ data in figure \ref{fig:ISTSAT_PrNby2_main_text}), compared to exponential decay in the uniform field case shown in \ref{app:taqc_as_a_seed}, with an exponent predicted analytically in \cite{kapit2024approximability}. Making this choice requires knowledge of $G$, and thus a solution to the problem, so it's not obvious that this discovery will help. 

The observation that supports IST-SAT is as follows: suppose the phases $\varphi_j$ are guessed correctly with some probability $r \geq 1/2$, based on the values of each bit in $G$. We note that the relative phases in $P$ are most crucial. As $r$ increases toward 1, the probabilities of finding any approximate solution nearest in Hamming distance monotonically increase at equivalent total evolution time, as shown in figure \ref{fig:ISTSAT_PrNby2_main_text}. Therefore, using the bitstrings from previous iterations of IST-SAT to construct the phase pattern for subsequent iterations yields solutions that are progressively closer to the ground state, enabling the algorithm to converge rapidly.

The results obtained from simulation in this work consistently present an empirical critical Hamming radius (CHR) $r_c$, which depends on $N_C/N$ and the unsatisfied fraction $\epsilon$ in the planted solution. We define the CHR $r_c$ in the following way: suppose each local phase is guessed correctly (relative to the planted solution) with probability $1-r$. Then, for linearly growing $t_f$, $r_c$ is defined to be the largest value of $r$ such that the probability of returning states with a Hamming distance $\leq r N$ is constant with increasing $N$. For $N_C/N = \cuof{1.5,2,4}$ and $\epsilon=0.1$, we respectively observe $r_c \simeq \cuof{1/4,1/4,1/3}$, which demonstrates that problems with larger constraint densities are solved with comparably worse initial approximations. In simulations of smaller systems with larger constraint densities (data not shown), the CHR are observed to be consistent with $r_c \simeq 1/3$.

Therefore, if the approximate seed strings are within some $r < r_c$, the bitstrings returned each round in IST-SAT will get closer and closer to the global optimum; with $r_c$ defined in this way, we expect convergence from a string less than $r_c N$ flips to the planted solution in a polynomial number of algorithm iterations. If we start from strings which are more than $r_c$ flips away, iterating phase patterns are still expected to converge to a global optima faster than the TAQC algorithm IST-SAT uses as a starting point. This is demonstrated by a smaller scaling exponent which can be inferred from the smaller slopes for curves with $r_c < r < 1/2$ in \ref{app:taqc_as_a_seed} as compared to the TAQC data, though the quantitative analysis of scaling is more complex in that case.

Throughout this paper, we define the time-to-solution (TTS) as the exponent $b$ obtained from exponential fits to numerical data presented in this work. We use the function $f(N) = a2^{bN}$, where $a$ is a pre-factor constant, and $b$ is the exponent which indicates the expected number of shots it takes to reach the a ground state of the problem. Hence, when $b \geq 0$, IST-SAT returns solutions scaling as a polynomial in $N$, while $b < 0$ implies an exponentially increasing with number of iterations to reach a global minima.

\begin{table}[H]
    \caption{\label{tab:TTS} Inferred time-to-solution (TTS) to reach a global optima for various algorithms: a semi-greedy classical (SGC) algorithm \cite{bellitti2021entropic}, TAQC \cite{farhi2014quantum}, SGC seeded with TAQC, and IST-SAT seeded with TAQC. The TTS (defined above in section \ref{sec:results}) is derived from the scaling exponent $b$ in numerical fits to simulation data.}
    \begin{indented}
    \item[]\begin{tabular}{@{}llll}
    \br
    Algorithm set-up & $N_C/N=1.5$ & $N_C/N=2$ & $N_C/N=4$ \\
    \mr
            SGC & -0.092 & -0.15 & -0.050\\
            TAQC & -0.11 & -0.17 & -0.20\\
            TAQC$\rightarrow$SGC & -0.073 & -0.081 & -0.025\\
            TAQC$\rightarrow$IST-SAT & -0.026 & -0.053 & -0.027 \\
    \br
    \end{tabular}
    \end{indented}
\end{table}

For the TAQC$\rightarrow$SGC sequence in table \ref{tab:TTS}, we report the combination with the least exponentially decaying exponent from \ref{tab:greedy}. For the TAQC$\rightarrow$IST-SAT sequence, we report the TAQC exponent from \ref{tab:xor_non_AC_fits} corresponding to the approximation distance equal to the critical Hamming radius ($d = r_c$) of IST-SAT. The full set of exponents for other distances $d \neq r_c$ may be found in tables in the referenced appendices. The inferred TTS demonstrates that IST-SAT obtains a significant polynomial speed-up over SGC and TAQC at all constraint densities. The performance of TAQC$\rightarrow$IST-SAT out-performs TAQC$\rightarrow$SGC at $N_C/N = \{1.5,2\}$, while demonstrating comparatively worse performance at $N_C/N = 4$. We attribute this difference due to the warm-started SGC algorithm not reaching expected asymptotic exponential decay at $N_C/N = 4$ until system sizes beyond our simulation capabilities $(N > 200)$, while TAQC and IST-SAT demonstrate the expected asymptotic scaling in small problem sizes. Furtherperformance results of the warm-started SGC algorithm may be found in \ref{app:warm_start_sgc}.

\begin{figure}[H]
    \centering
    \includegraphics[width=\textwidth]{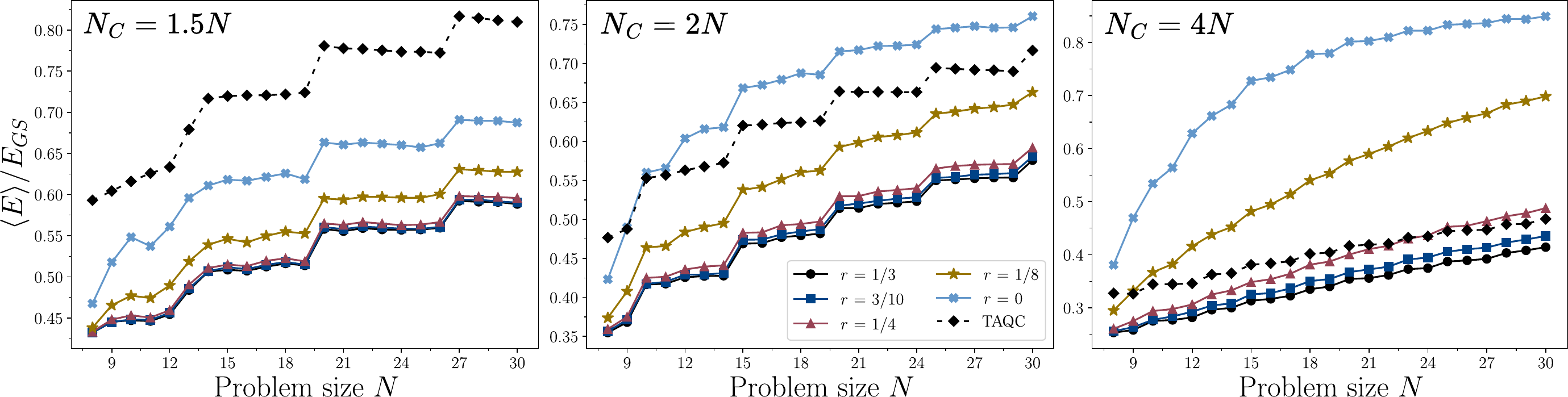}
    \caption{Average energy $\langle E \rangle $ returned from IST-SAT with a lower frequency drive $\omega 10\pi \ln N$ (\ref{app:istsat_low_freq}) and TAQC, normalized by the ground state energy $E_{GS}$. Energy is obtained using $H_P$ as the cost function. Here, we run a single iteration of IST-SAT, and use 10,000 samples (bitstrings) from the wave function in exact state vector simulation. The legend in the center figure specifies the fraction of incorrect bits $rN$ used in the seed state protocol.}
    \label{ist_qaoa_energy}
\end{figure}

To examine the \emph{quality of strings} returned from IST-SAT, we plot the average energy $\langle E \rangle$ (normalized by $E_{GS}$) in figure \ref{ist_qaoa_energy}. We note that the data reported in figure \ref{ist_qaoa_energy} were obtained using a lower frequency of $\omega = 10 \pi \ln N$, compared to the data reported in figure \ref{fig:ISTSAT_PrNby2_main_text}. Interestingly, the rapid convergence to any global optima is based only on Hamming distance $D_{\mathrm{H}}$ and remains uncorrelated with the energies $E$ of the returned states. However, we still observe monotonically increasing average string qualities as IST-SAT is seeded with better phase patterns. Our very first formulation of IST-SAT used a variation of the quasi-greedy algorithm in \cite{bellitti2021entropic} to gather the initial seed states, and we observed no improvement over random guessing in making this choice for reasons discussed below. This is supported by the result that, unless $r$ is very close to one, the average energy returned by IST-SAT is somewhat worse than that of TAQC for all other parameters equal (see the supplemental information for figures). We attribute this observation to there being many local minima with energies close to $E_{GS}$, and the effect of the high frequency oscillating drive is to steer the evolving state away from them and toward $\ket{G}$ and its local excitations, which may be higher in energy even if they are much closer in Hamming distance.

While naive classical seeding was unsuccessful, matching the observed $r_c \simeq \cuof{1/4,1/4,1/3}$ with the corresponding distance probabilities for TAQC yields significantly reduced asymptotic time to solution if strings from TAQC (or equivalently, random phase patterns) are used for initial seeding. Selecting fractions of the TAQC strings based on their energies yielded no benefit over randomly sampling the TAQC output distribution.

Interestingly, the results demonstrate that even when the initial seed string is chosen relative to a single solution (the planted solution), the probability of reaching other equally valid solutions remains robust. This result is clearly observed at the smallest constraint density $(N_C/N = 1.5)$, where each problem instance tends to have several global optima. At larger constraint densities of $N_C/N = \{2,4\}$, the probability of reaching any solution more closely matches the probability of reaching the planted solution, which is unique with higher probability. In \ref{app:further_istsat_details}, we discuss the choice of phase pattern, convergence to the planted solution in particular, and the exact solution statistics of the problem instances used in this work.

\section{Conclusion}

We have introduced IST-SAT, a quantum algorithm that uses a non-classical steering mechanism based on high frequency oscillating drives that guides the algorithm towards the global minima of hard optimization problems. IST-SAT is a unique feedback-based quantum algorithm which does not require the calculation of any gradients or averages used in the original variational QAOA proposal, or more recent versions such as recursive QAOA \cite{bravyi2022hybrid}. These costly methods require hundreds to thousands of shots to calculate accurately. The core mechanism for how the phase pattern appropriately guides the optimization is only understood by analogy to other work in a simpler system \cite{grattan2024exponential}, where we expect a careful analytical derivation of the speedups observed in this work to inform further algorithm innovations. 

Regarding seed algorithms for IST-SAT, it may be the case that other classical or quantum algorithms could provide further benefits to for finding good initial approximations. For example, an algorithm that breaks $H_\mathrm{P}$ into exactly solvable sub-problems may obtain sets of relative phase patterns which potentially yield significant improvements. Future work may also test versions of IST-SAT by assigning each group of spins a different frequency (not just a different phase pattern), for better averaging, as in the earlier multi-frequency AC optimization schemes \cite{kapit2021systems,kapit2021noise,tang2021unconventional,mossi2023embedding} that inspired this work. 

While all phase patterns were set using discrete offsets of $\{0, \pi\}$ in this work, we expect that more sophisticated phase pattern selection protocols would further improve IST-SAT. An interesting future direction could consider using semi-definite programs to selecting phases in the $H_{\mathrm{ST}}$ drive in a continuous interval $\varphi_j = \left[0,\pi \right]$, building off recent improvements to a variational quantum algorithm \cite{King2023improved} for quantum Max-Cut. We expect that the core mechanism of IST-SAT is broadly applicable to other problems in optimization, such as Max-Cut or low auto-correlated binary sequences (LABS), which we leave for future exploration. Finally, it is expected that this IST-SAT is not single-error fragile, in contrast to recently proposed quantum algorithms for MAX-3-XORSAT \cite{kapit2024approximability}, thus motivating future experiments on near-term quantum devices.

\ack
We would like to thank Joao Basso and Ojas Parekh for insightful discussions. We would also like to thank Takuto Komatsuki and Joey Liu for support with quantum simulations on the Fujitsu Quantum Simulator. This work was supported by the DARPA Reversible Quantum Machine Learning and Simulation pro- gram under contract HR00112190068, as well as by National Science Foundation grants PHY-1653820, PHY-2210566, DGE-2125899, and by the U.S. Department of Energy, Office of Science, National Quantum Information Science Research Centers, Superconducting Quantum Materials and Systems Center (SQMS) under contract number DE-AC02-07CH11359. The SQMS Center supported EK’s advisory role in this project, as well as his time improving and fine tuning the algorithms. Part of this research was performed while BB was visiting the Institute for Pure and Applied Mathematics (IPAM), which is supported by the National Science Foundation (Grant No. DMS-1925919). This work was performed in part at Aspen Center for Physics, which is supported by National Science Foundation grant PHY-2210452. The Flatiron Institute is a division of the Simons Foundation.

$^{\dagger}$BB and JS contributed equally to this work.

\appendix

\section*{Appendix}

\section{Further details on performance of IST-SAT}
\label{app:further_istsat_details}

\subsection{Phase pattern selection and convergence to the planted solution}
\begin{figure}[H]
    \centering
    \includegraphics[width=\textwidth]{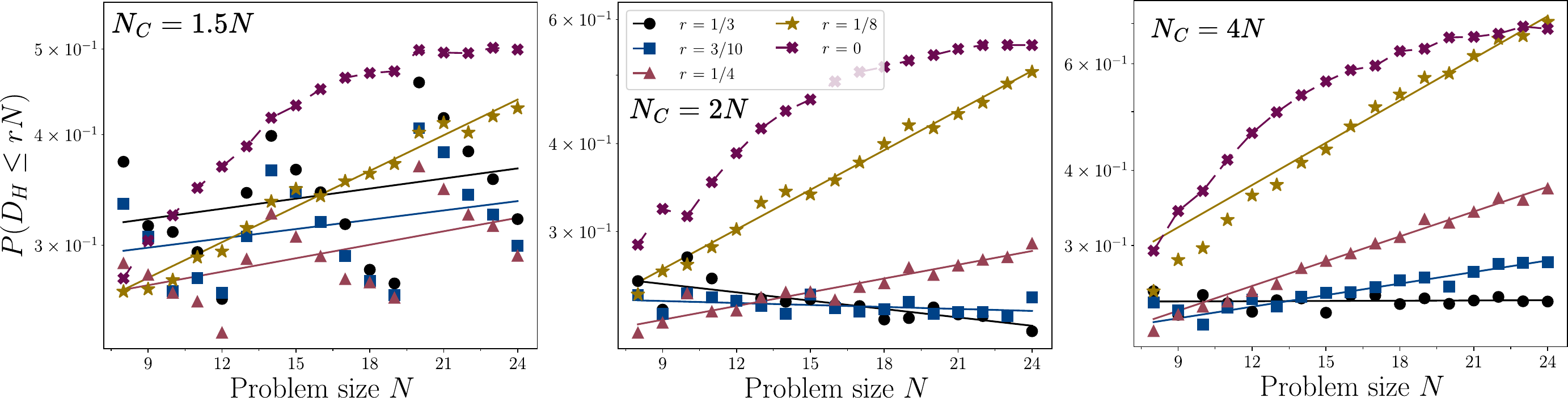}
    \newline \newline\includegraphics[width=\textwidth]{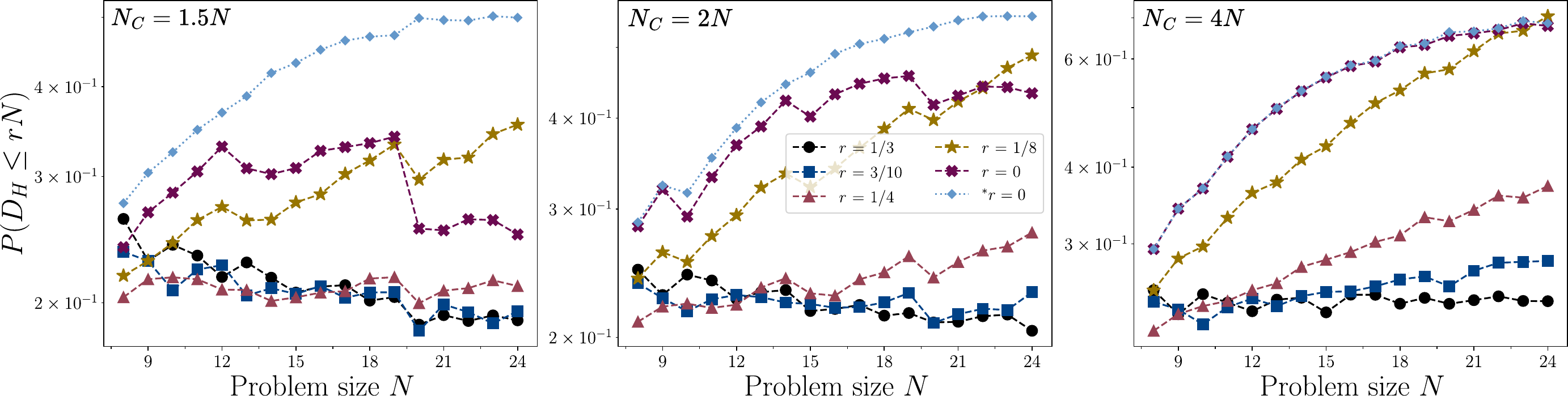}
    \caption{Convergence of IST-SAT to the (top row) nearest solution in Hamming distance and (bottom row) the planted solution. Initial phase patterns $P_0$ were chosen from random seed strings with $(1-r)$ of the bits correctly chosen relative to the planted solution. We include the $r=0$ case from the top row as the $r^* = 0$ case in the bottom row (blue diamonds) to show how IST-SAT targets the planted solution at high constraint densities.}
    \label{fig:ISTSAT_pattern_selection}
\end{figure}

We report the convergence of IST-SAT to the nearest solution in Hamming distance and the planted solution in particular shown in figure \ref{fig:ISTSAT_pattern_selection}. Notably, the results for convergence to any solution at $N_C/N=1.5$ demonstrate a positive exponent for all fractions $r$ of the random approximate seed string. All exponential fits for the top row of figure \ref{fig:ISTSAT_pattern_selection} can be found in \ref{tab:any_GS_fits}. When considering the convergence to the planted solution in particular, the performance of IST-SAT notably degrades at $r=0$. W attribute this behavior to the problem instances containing several global minima (multiple ground states) at lower constraint density, which is supported by exact solution statistics provided in \ref{app:exact_solution_stats}. As discussed in the main text, while performance of reaching the planted solution degrades, IST-SAT remains robust to finding other global optima equal in energy to the planted solution.

While low constraint densities have more ground states, the number of solutions is typically very small at larger constraint density $(N_C/N =4)$, where the convergence to any solution and the convergence to the planted solution are nearly identical (see right column of figure \ref{fig:ISTSAT_pattern_selection}). This alignment in convergence suggests that IST-SAT targets not only the solution to which the pattern was pre-selected for, but all equally valid solutions. Therefore, in the problem instances we tested in this work, a pattern selected relative to one solution remains robust to finding other global minima. These results are supported by exact solution statistics in \ref{app:exact_solution_stats} for the MAX-3-XORSAT problem instances we consider in this work.

\subsection{Performance of IST-SAT with random approximate seed strings}
\label{app:seeded_istsat}
\begin{table}[H]
    \label{tab:any_GS_fits}
    \caption{Exponential fits to the numerical data presented for IST-SAT seeded with random approximate seed strings for probabilities $P(D_{\mathrm{H}} \leq rN)$ (see top row of figure \ref{fig:ISTSAT_pattern_selection}) and $P(D_{\mathrm{H}} \leq rN/2)$ (see figure \ref{fig:ISTSAT_PrNby2_main_text}). In this table, we report the exponent $b$ from fitting the numerical data to $a2^{bN}$. The exponents corresponding to the critical Hamming radius are highlighted in bold text.}
    \begin{indented}
        \item (a)
        \item[]\begin{tabular}{@{}llll}
        \br
             Guessing error $r$ & $N_C/N=1.5$ & $N_C/N=2$ & $N_C/N=4$ \\
        \mr
            $1/3$ & 0.013 & -0.013 & \textbf{0.00045} \\ 
            $3/10$ & 0.012 & -0.0031  & 0.021 \\ 
            $1/4$ & \textbf{0.017} & \textbf{0.022} & 0.046 \\ 
            $1/8$ & 0.045 & 0.062 & 0.077 \\ 
        \br
        \end{tabular}
        \item (b)
        \item[]\begin{tabular}{@{}llll}
        \br
             Guessing error $r$ & $N_C/N=1.5$ & $N_C/N=2$ & $N_C/N=4$ \\
        \mr
            $1/3$ & -0.024 & -0.038 & \textbf{0.0013} \\ 
            $3/10$ & -0.016 & -0.020 & 0.028 \\ 
            $1/4$ & \textbf{-0.00087} & \textbf{0.012} & 0.056 \\ 
            $1/8$ & 0.041 & 0.060 & 0.084 \\ 
        \br
        \end{tabular}
    \end{indented}
\end{table}

\subsection{Exact solution statistics at small problem sizes}
\label{app:exact_solution_stats}
\begin{figure}[H]
    \centering
    \includegraphics[width=\textwidth]{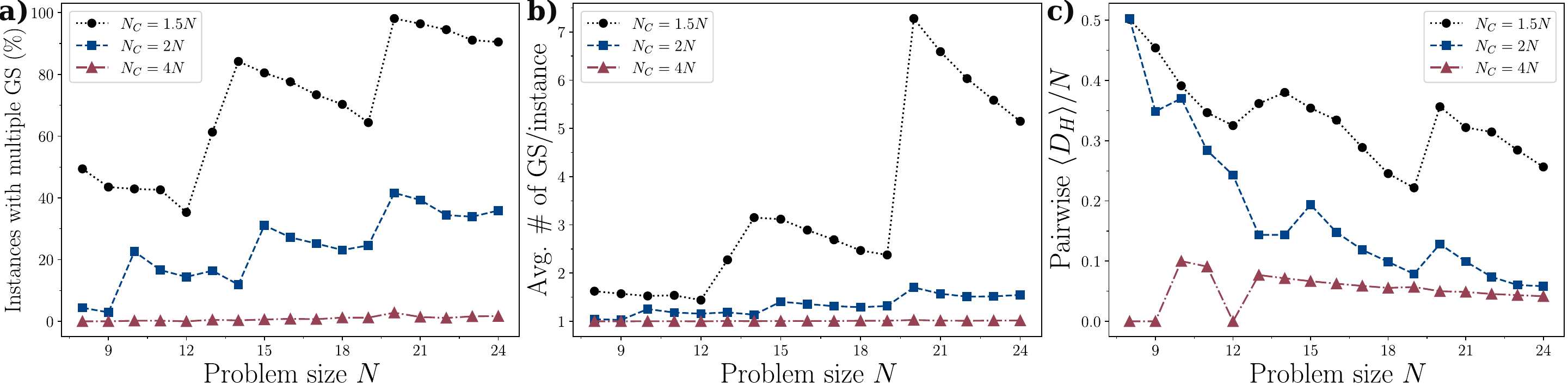}
    \caption{(a) Percentage of problem instances (from the total 1000 instances) with more than one ground state, including the PPS. For each system size $N=8-24$, we computed exact ground states using brute-force methods. (b) Average number of ground states per problem instance. (c) Pairwise Hamming distance $D_{\mathrm{H}}$ within the set of ground states. The $N=8,9,12$ cases have only one ground state, which is the PPS.}
    \label{fig:multiple_gs}
\end{figure}

To produce figure \ref{fig:multiple_gs}, we calculated the exact number of exact solutions via brute force search to $N=24$, for 1000 problem instances at each system size. In figure \ref{fig:multiple_gs} (a-b), the problem instances at $N_C/N=\{1.5,2\}$ tend to have more than one ground state, while at $N_C/N=4$, the planted solution tends to be a unique ground state. These statistics support the results shown in figure \ref{fig:ISTSAT_pattern_selection}, where the convergence to the any solution does not show any exponential decay for all approximate seed state distances. For the largest constraint density we present in this work ($N_C/N=4$), the planted solution is nearly always the unique global optima. 

When a given instance has multiple ground states, we further would like to understand how these solutions are correlated. In figure \ref{fig:multiple_gs} (c) we report the pairwise Hamming distance between ground states, which demonstrates that even when a problem contains several solutions, the solutions tend to very close to each other measured in Hamming distance. This strong correlation demonstrates that solutions tend to be only a few bit flips away from each other on average. These statistics support the claim that when the phase patterns are selected only relative to one solution (the planted solution, which is known), IST-SAT is robust to finding other global minima due a small pairwise Hamming distance withing the set of optimal solutions. Future work may consider other problems with instances in which there are no such strong correlations are expected.

\section{Seeding with trotterized adiabatic quantum computation}
\label{app:taqc_as_a_seed}

\begin{figure}[H]
\includegraphics[width=\textwidth]{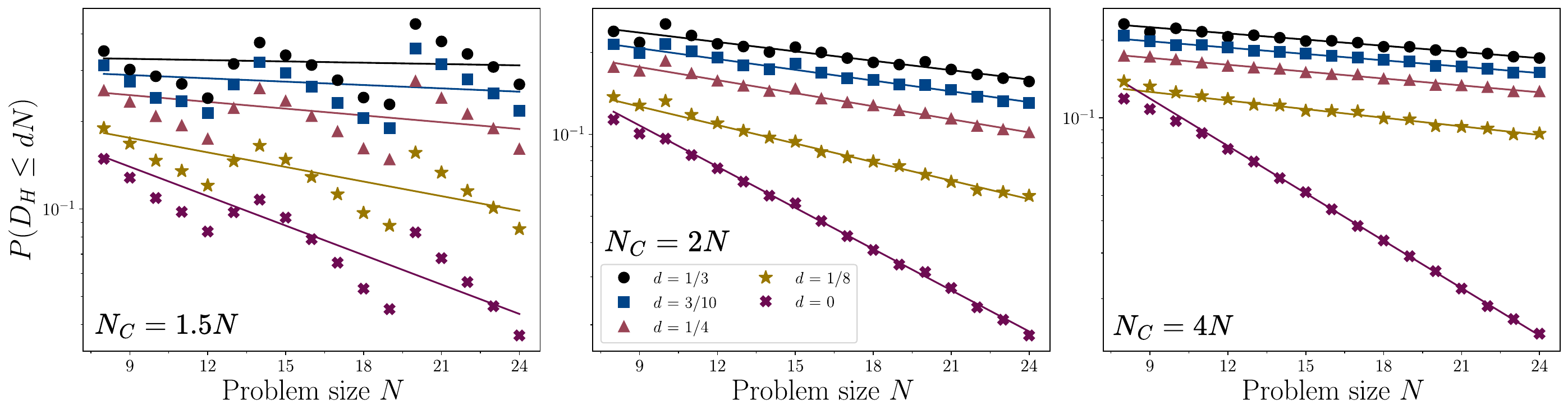}
\caption{Probabilities of TAQC returning approximate strings within a Hamming distance $D_{\mathrm{H}} \leq d N$ from the nearest solution for (left to right) $N_C/N = \cuof{1.5,2,4}$. Data are fit to lines $a 2^{b N}$, where the exponents $b$ are reported in \ref{tab:xor_non_AC_fits}.}
\label{fig:qaoa_PdN}
\end{figure}

In this section, we consider the capabilities of TAQC as a seed algorithm for IST-SAT. Specifically, we report the approximation performance of the TAQC and SGC algorithms in terms of Hamming distance to any ground state. In figure \ref{fig:qaoa_PdN}, we show the probabilities of reaching approximate solutions, measured by an approximation in Hamming distance $D_{\mathrm{H}} \leq dN$ to the nearest solution, where $d$ is some fraction in the range $\left[0, 1/2\right]$. Exact solutions are measured by $d=0$, while $d=1/2$ is equivalent to random guessing, and $0 < d < 1/2$ are approximate solutions.

\begin{table}[H]\label{tab:xor_non_AC_fits}
    \caption{Exponential fits to the numerical data presented for TAQC in figure \ref{fig:qaoa_PdN}. In this table, we report the parameters $b$ obtained from fitting the function $a2^{bN}$ to the probabilities $P(D_{\mathrm{H}} \leq dN)$ of finding states close in Hamming distance $D_{\mathrm{H}}$ for different distances $d N$ or fewer flips away from the nearest solution. We include the fitting results for the set of constraint densities $N_C/N = \{1.5, 2, 4\}$. We denote the fraction $d$ for each constraint density associated to $r_c$ in bold face text.}
    \begin{indented}
        \item[]\begin{tabular}{@{}llll}
        \br
        Fractional Hamming distance $d$ & $N_C/N=1.5$ & $N_C/N=2$ & $N_C/N=4$ \\
        \mr
            $1/3$ & -0.0050 & -0.038 & \textbf{-0.027}\\ 
            $3/10$ & -0.013 & -0.044 & -0.027\\ 
            $1/4$& \textbf{-0.026} & \textbf{-0.053} & -0.030\\
            $1/8$ & -0.056 & -0.075 & -0.037\\
            $0 (P_{\mathrm{GS}})$ & -0.11 & -0.17 & -0.20\\ 
        \br
        \end{tabular}
    \end{indented}
\end{table}

\section{Performance of the semi-greedy classical algorithm}
\label{app:sgc_performance}

\begin{figure}[H]
\includegraphics[width=\textwidth]{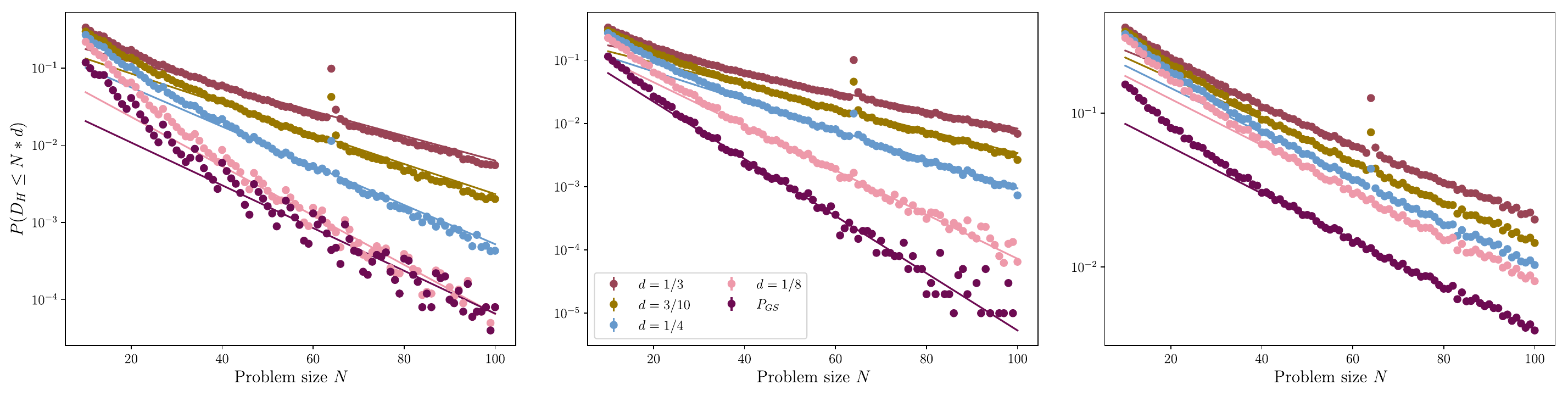}
\caption{Probabilities of SGC returning approximate strings within a Hamming distance $D_{\mathrm{H}} \leq d N$ from the \emph{planted solution} for (left to right) $N_C/N = \cuof{1.5,2,4}$. Data is plotted as $P(D_{\mathrm{H}} \leq d N) + P_{\mathrm{GS}}$ to estimate the probability of finding any ground state. Data are fit to lines $a 2^{b N}$, where the exponents $b$ are reported in \ref{tab:sgc_pdn_exponents}.}
\label{fig:sgc_PdN}
\end{figure}

\begin{table}[H]\label{tab:sgc_pdn_exponents}
    \caption{Exponential fits to $a2^{bN}$ for the SGC data presented in figure \ref{fig:sgc_PdN}.}
    \begin{indented}
        \item[]\begin{tabular}{@{}llll}
        \br
        Fractional Hamming distance $d$ & $N_C/N=1.5$ & $N_C/N=2$ & $N_C/N=4$ \\
        \mr
            $1/3$ & -0.053 & -0.049 & -0.040 \\ 
            $3/10$ & -0.065 & -0.059 & -0.044 \\
            $1/4$ & -0.084 & -0.077 & -0.048 \\ 
            $1/8$ & -0.11 & -0.12 & -0.050 \\ 
            $0 (P_{\mathrm{GS}})$ & -0.092 & -0.15 & -0.050 \\ 
        \br
        \end{tabular}
    \end{indented}
\end{table}

\section{Warm starting the semi-greedy classical algorithm}
\label{app:warm_start_sgc}

To further approximate the critical Hamming radius for our PPSPs we applied a classical semi-greedy descent to random problem instances with constraint densities $N_C/N = \{1.5,2,4\}$. The algorithm is a modified version of the simple greedy algorithm introduced in \cite{bellitti2021entropic} and later applied to hypergraphs in \cite{kapit2024approximability}. To approximate the critical Hamming radius, we ran the semi-greedy algorithm on random hypergraphs for the stated constraint densities. The initial states started from distances $rN$ flips away from the planted solution to show the SGC algorithm performance when starting from approximate solutions nearby the planted solution. 

\begin{figure}[H]
    \centering
    \includegraphics[width=\textwidth]{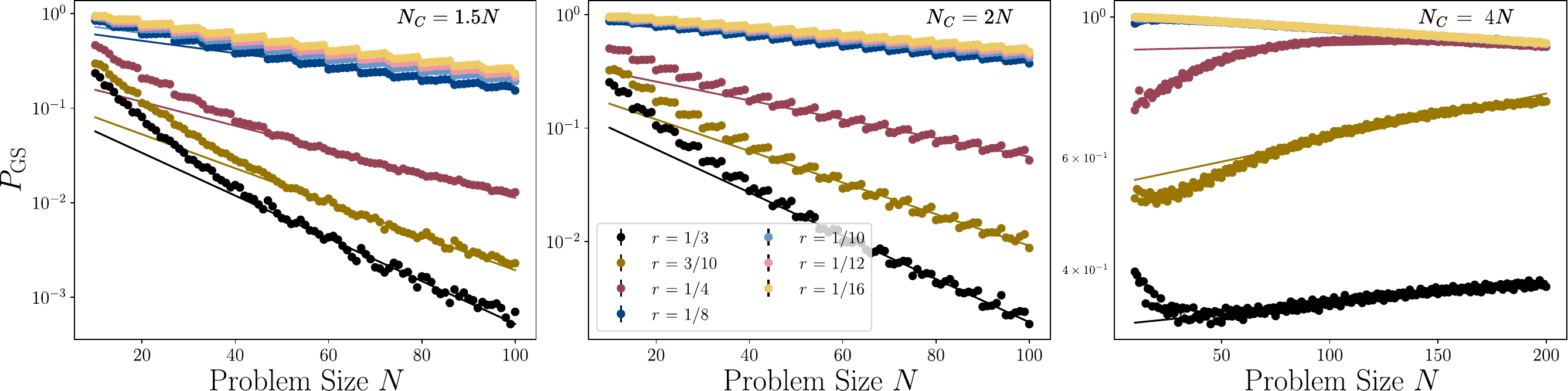}
    \caption{Probabilities of finding the ground state of random hypergraphs for $N_{C}/N=\{1.5, 2, 4\}$ using 100,000 trials on each constraint density and each value of $rN$. The data was fit to $a2^{b N}$ using problem sizes from $N=40$ to $N=100$. The results for the $4N$ constraint density demonstrated unusual behaviors at smaller $N$, so we ran it to $N=200$ to try and find convergent behavior. The fits were calculated on the data points from $N=100$ to $N=200$. We found that the seeding the algorithm from farther away resulted in an increasing probability in finding the ground state with $N$, however it is clear that these trends eventually become an exponential decay in finding the ground state. With all of that considered, the algorithm still performs best when seeded with values close to the ground state.}
    \label{Greedy}
\end{figure}

\begin{table}[H]\label{tab:greedy}
    \caption{In (a), we report the exponential fits to the warm started SGC algorithm for various guessing errors $r \in [1/16, 1/3]$ shown in figure \ref{Greedy}. In (b), we report the inferred performance of the algorithm set-up: TAQC$\rightarrow$SGC, where TAQC is used to produce a string to warm start the SGC algorithm. The exponents in (b) are calculating by adding the exponent from TAQC in \ref{tab:xor_non_AC_fits} to the exponents in (a), at fractions $d = r$, where $d$ is the approximation distance of TAQC, and $r$ is the guessing error for the warm start of the SGC algorithm. We highlight the best combination of exponents in (b) using bold text, which are reported in table \ref{tab:TTS} in the main text.}
    \begin{indented}
        \item (a)
        \item[]\begin{tabular}{@{}llll}
        \br
        Guessing error $r$ & $N_C/N=1.5$ & $N_C/N=2$ & $N_C/N=4$ \\
        \mr
                $1/3$ & -0.075 & -0.063 & 0.00093 \\ 
                $3/10$ & -0.060 & -0.046 & 0.0017 \\ 
                $1/4$ & -0.042 & -0.028 & -0.00040 \\ 
                $1/8$ & -0.022 & -0.013 & -0.00077 \\ 
                $1/10$ & -0.021 & -0.012 & -0.00076 \\ 
                $1/12$ & -0.021 & -0.012 & -0.00076 \\ 
                $1/16$ & -0.021 & -0.011 & -0.00075 \\ 
        \br
        \end{tabular}
        \item (b)
        \item[]\begin{tabular}{@{}llll}
        \br
        Guessing error $r$ & $N_C/N=1.5$ & $N_C/N=2$ & $N_C/N=4$ \\
        \mr
                $1/3$ & -0.080 & -0.10 & -0.026 \\ 
                $3/10$ & \textbf{-0.073} & -0.09 & \textbf{-0.025}\\ 
                $1/4$ & -0.068 & \textbf{-0.081} & -0.030\\ 
                $1/8$ & -0.078  & -0.088 & -0.038\\ 
        \br
        \end{tabular}
    \end{indented}
\end{table}

\section{Performance of IST-SAT with low frequency AC drives}
\label{app:istsat_low_freq}

In this section, we report the results obtained for IST-SAT with lower frequency of $\omega=10\pi\ln N$ and shorter run-time that was simulated to larger system size $(N=30)$. While this parameter set under performs higher frequency drives and with longer run-times as presented in the main text, the empirical radii of convergence we identify do not change significantly, except at $N_C/N = 1.5$, where the difference is performance is most significant. One may consider a finer-grid search to identify how exactly $r_c$ scales with the constraint density $N_C/N$ and fraction $\epsilon$ of unsatisfied constraints in the ground state.

\begin{figure}[H]
    \includegraphics[width=\textwidth]{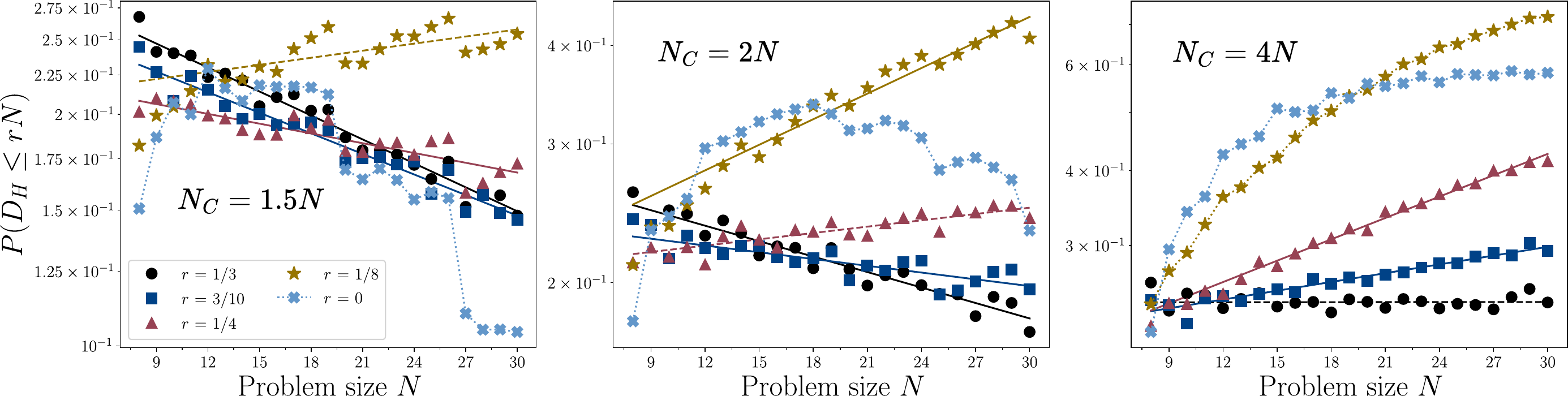}
    \newline \newline
    \includegraphics[width=\textwidth]{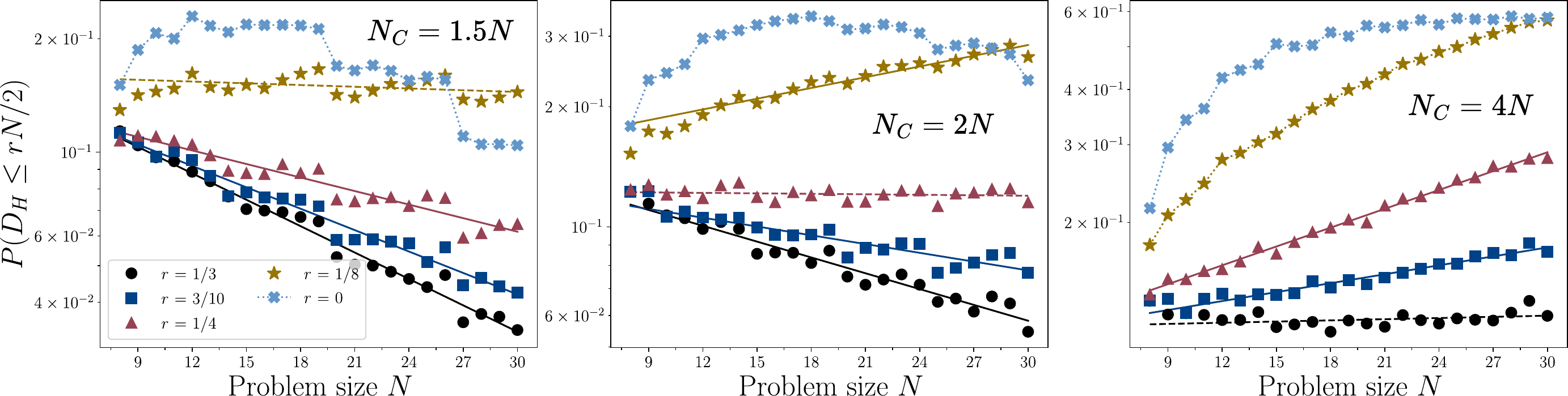}
    \caption{Convergence of IST-SAT to the planted solution. Probabilities of returning a solution with probabilities $P(D_{\mathrm{H}} \leq rN)$ and $P(D_{\mathrm{H}} \leq rN/2)$ are presented in the top and bottom row respectively. A critical Hamming radius $r_c$ can be readily identified in these plots by dashed lines, which are approximately the first fraction $r$ to obtain constant probability with increasing $N$.}
    \label{ACrc}
\end{figure}

Shown in figure \ref{ACrc}, we observe monotonically increasing probabilities of returning solutions that approximate the PPS at different approximation ratios $rN$ in terms of Hamming distance $D_{\mathrm{H}}$. We identify a critical Hamming radius $r_c$ for each constraint density $N_C/N = \{1.5, 2, 4\}$ respectively as $r_c = \{1/8,1/4,1/3\}$.

\begin{table}[H]\label{tab:IST-SAT_fits}
    \caption{Exponential fits to the low frequency IST-SAT data presented in figure \ref{ACrc}. We report the exponents $b$ from fitting data to $a2^{bN}$ where (a) is $P(D_{\mathrm{H}} \leq rN)$ data, and (b) is the fit to $P(D_{\mathrm{H}} \leq rN/2)$ data. The exponents lying at the approximate radii of convergence $r_c = \{1/8,1/4,1/3\}$ for $N_c/N=\{1.5,2,4\}$ (respectively) are highlighted in bold text. The $--$ entries represent data where exponential fits could not accurately represent the data.}
    \begin{indented}
        \item (a)
        \item[]\begin{tabular}{@{}llll}
        \br
             Guessing error $r$ & $N_C/N=1.5$ & $N_C/N=2$ & $N_C/N=4$ \\
        \mr
                $1/3$ & 0.034 & 0.022 & \textbf{-0.000083}\\
                $3/10$ & 0.030 & 0.00095 & -0.016\\
                $1/4$ & 0.014 & \textbf{-0.0089} & -0.039 \\
                $1/8$ & \textbf{-0.010} & -0.036 & --\\
        \br
        \end{tabular}
        \item (b)
        \item[]\begin{tabular}{@{}llll}
        \br
             Guessing error $r$ & $N_C/N=1.5$ & $N_C/N=2$ & $N_C/N=4$ \\
        \mr
            $1/3$ & 0.078 & 0.044 & \textbf{-0.030}  \\ 
            $3/10$ & 0.063 & 0.024 & -0.022  \\ 
            $1/4$ & 0.040 & \textbf{0.0013} & -0.047  \\ 
            $1/8$ & \textbf{0.0051} & -0.030 & --  \\ 
        \br
        \end{tabular}
    \end{indented}
\end{table}

\clearpage

\section*{References}
\bibliographystyle{unsrt}
\bibliography{main}

\clearpage

\end{document}